\documentclass[a4]{article}
\usepackage{amsmath}

\begin{document}

\title{Geometric unification theory of the grand unification and gravitational interactions and new physics}
\author{C. Huang$^{1,2,*}$ and Yong-Chang Huang$^{3,4,+}$ \\
%EndAName
\\
$^1$Lawrence Berkeley National Laboratory, 1 Cyclotron Road, Berkeley CA
94720, USA\\
$^2$Department of Physics and Astronomy, Purdue University, \\
525 Northwestern Avenue, W. Lafayette, IN 47907-2036, USA\\
$^3$Institute of Theoretical Physics, Beijing University of Technology, \\
Beijing 100124, China\\
$^4$Institute of Theoretical Physics, Jiangxi Normal University, \\
Nanchang, 330022, China\\
\\
$*$email:c.huang0@hotmail.com: +email:ychuang@bjut.edu.cn \\
}
\maketitle

\begin{abstract}
\bigskip This paper discovers geometric unification theory of the grand unification and gravitational interactions and their new physics
according to the general fiber bundle theory, symmetry and so on.
Consequently, the research of this paper is based on the exact scientific
bases of mathematics and physics. The Lagrangians of the grand unification and gravitational interactions are unifiedly deduced from quantitative causal
principle (QCP) and satisfy the gauge invariant principle of general gauge
fields interacting with Fermion and/or boson fields. The geometry and
physics meanings of gauge invariant property of different physical systems
are revealed, and it is discovered that all the Lagrangians of the four
fundamental physics interactions are composed of the invariant quantities in
corresponding spacetime structures. The difficulties that fundamental
physics interactions and Noether theorem are not able to be unifiedly given
and investigated are overcome, the geometric unification theory and origins
of the four fundamental physics interactions and Noether theorem are shown
by QCP, their two-order general Euler-Lagrange Equations and corresponding
Noether conservation currents are derived in general curved spacetime. This
paper further deduces QCP from symmetric principle. Consequently, geometric
unification theory of the grand unification and gravitation theories and
Noether theorem based on symmetric principle and the new physics are given
in this paper. This paper further gives the unification of QCP and symmetric
principle. Thus, this paper opens a door to both study and give new
developments of geometric unification theory of physics laws, and using the
new geometric unification theory, a lot of research works about different
branches of physics can be anew done and expressed simpler with different
symmetric characters.

%EndAName

%EndAName
\end{abstract}

\section{Introduction}

Causal principle should be satisfied in expressing physical laws. In quantum
field theory, the causal principle is reflected in the results that if the
square of the distance of spacetime coordinates of two boson (or fermion)
operators is timelike, the canonical commutator (or anticommutator) of the
boson (or fermion) field does not vanish, i.e., their measures are coherent,
but not coherent for spacelike case \cite{wen}. Dispersion relations are
deduced by means of the causal principle etc \cite{kle}. Ref. \cite{cur}
studied the relationships of causal principle and symmetry principle.

It is well known that in physics the well-admitted physical law not
satisfying causal principle has not been found. In terms of the condition
that causal principle is quantitative, i.e., quantitative causal principle (
QCP ), Ref\textsl{. }\cite{hdf} gives a unified theory of all differential
variational principles, and a unified theory of all integral variational
principles are achieved \cite{htt}. The investigations about the
generalization from classical statistical mechanics to quantum mechanics
still satisfy quantitative causal principle\ \cite{ab}. Refs. \cite%
{hdf,htt,hin} show the quantitative causal principle of gain, loss and
transformation of any system, and prove that the equivalent causes must
result in the equivalent results, which is just the invariant property of
some operations, namely, symmetry properties \cite{uni,wei}. In fact, the
symmetric properties in physics and mathematics are just the invariant
properties under some kinds of operations of systems \cite{wei}, almost all
physics processes, e.g., Refs.\cite{wu,wu1}, satisfy causal relations
because causal relations should be satisfied in the universe, the relation
between symmetries and QCP is given \cite{hdf,htt}, and using QCP, Ref. \cite%
{uni} gives\ unification theory of different causal algebras and its
applications to theoretical physics, in which group theory is included as a
special subset.

Utilizing the no-loss-no-gain homeomorphic map transformation satisfying
QCP, Ref. \cite{hin} overcomes the non-perfect properties of the Volterra
process, gains the exact strain tensor formulae in condensed theory, and
solves the hard difficulty to give the really serious one-to-one
correspondences between the two theories of dislocations in Euclidean space
and Weitzenb\"{o}ck Material Manifolds. Ref. \cite{htt} gives the unified
origin of different variational principle and Noether theorem in terms of
QCP. Now it is well known that Noether theorem is deduced from variational
principle and group theory \cite{noe,dju,liz}, while variational principle
and Noether theorem may be derived by QCP \cite{htt}. It can be seen that
this paper and Refs. \cite{hdf,htt,hin} are essential developments of the
investigations of Ref. \cite{cur} about the relationships of causal
principle and symmetry principle.\ It is well known that physical
fundamental interactions and Noether theorem are used to be regarded as
independent fundamental laws being not able to be unifiedly proved and
studied up to now, but in the following investigations we show that the two
laws are just the deductions of QCP and give the unified description of the
two laws.

Many Refs. \cite{Buchb,Henne,Buchbi,Rham,Caman} studied causal field
propagation that led to neat results, and these results are interesting and
also satisfy QCP. Unitarity and causality are key physical concepts in
physics studies, e.g., studying weak gravity conjecture \cite{weak}

In fact, superstring theory has gotten many important developments \cite%
{bbs,gary,Mir}, and Ref. \cite{gary1} very well investigated Quantum
modification of general relativity and so on. But it is well known that
there are still some key problems in various kinds of the grand or
supersymmetric unification theories up to now, e.g., there are too many
adjustable free parameters etc in various kinds of the grand or
supersymmetric unification theories, which are not natural, even superstring
theories and M-theory have themselves unsolved key problems, e.g., we still
cannot seriously break the high dimensional theories down to four dimensions
and do rigorous phenomenology etc. There are different unification theories,
e.g., from up to down (superstring) or from down to up (phenomenological)
unification theories, unification theory of this paper belongs to from down
to up unification theory. Therefore, exploring the other possible
unification theories is actually needed, which may activate to understand
and solve all the problems finally, that is, we may try to solve all the
problems from different aspects, which may finally activate development and
perfection of the various kinds of the unification theories, e.g.,
superstring theories, M-theory and so on.

Famous physical Nobel Prize winner S. Weinberg tells us \cite{wein}: The
present generation of young physicists may envy those of us who had the
excitement and delight of developing the standard model. This might be a
mistake, just as it turned out that my generation would have been mistaken
to envy the earlier heroes of quantum electrodynamics. Our newly minted
experimentalists and theorists now have a chance to participate in making
the next big step beyond the standard model. They may even be able to see
their way clear to the very high energy scale where a final theory will be
revealed. Thus we need to try various explorations.

The arrangement of this paper is: Sect. 2 is geometric unification theory of
fundamental physics interactions, Sect. 3 is the unified description of
fundamental physics interactions and Noether theorem, Sect. 4 is
applications of this theory and new physics, and Sect.5 is summary and
conclusions.

\section{Geometric Unification Theory of Fundamental Physics Interactions}

For the convenience of research, we simply review the proof of QCP. In
physics, quantitative action (cause) of some quantities must lead to the
equal action (result), i.e., how much lose (cause), how much gain (result),
which is just QCP deduced from the no-loss-no-gain principle in the universe
\cite{hdf,htt} and it may be concretely expressed as

\begin{equation}
DS-CS=0  \tag{1}
\end{equation}%
Eq.(1) means that any quantitative action produced by operator set D acting
on S must lead to appearance of set C acting on S so that DS equate CS,
where D and C may be different operator sets, the whole process satisfies
QCP so that right hand side of Eq.(1)\ keeps zero, i.e. satisfies the
no-loss-no-gain principle in the universe \cite{htt}. Thus, Eq.(1) is viewed
as a mathematical expression of QCP. Eq.(1) is very useful for the following
studies. Eq.(1)'s nonlinear expression refers to Ref. \cite{uni}.

All physics laws must satisfy causal principle and the causal principle must
be quantitative. Furthermore, up to now, all the physics laws violating
causal principle have not been found, it can be seen, using the quantitative
causal principle we shall be able to give the unification description
theory. Consequently, the principle is useful and important.

For a 4-dimensional physical general curved spacetime M$^{4}$ (\textsl{\
there naturally may exist the constructions of vector bundle }$E(M^{4},F,\pi
,G)$\textsl{\ or associate vector bundle }$E(M^{4},F,\pi
_{_{V}},G,P(M^{4},G,\pi ))$\textsl{\ of principal bundle }$P(M^{4},G,\pi )$%
\textsl{.\ Readers, not familiar with manifold and fibre bundle, don't need
to read the Italic type parts in parenthesis, which will not affect their
understanding on the paper, as the same below\ ), }taking $S_{V}$ as a
general basic vector field in any open neighborhood U$_{V}$ in M$^{4}$, D as
differential connection operator, C as connection $\omega _{V}$ in Eq.(1),
it follows that \cite{che,NS}
\begin{equation}
DS_{V}=\omega _{V}S_{V}  \tag{2}
\end{equation}%
Eq.(2)'s physical meaning is that the quantitative physical effect operator
D's acting on $S_{V}$ must equate that the gauge field or connection $\omega
_{V}$ times $S_{V}$ (\textsl{\ in which }$S_{V}$\textsl{\ may be a basic
vector of principal bundle whose base manifold is M}$^{4}$\textsl{\ or a
basic vector on M}$^{4}$\textsl{\ of vector bundle }).

Substituting the transformation $S_{V}=A_{VU}S_{U}$ into Eq.(2), it follows
that \cite{che}

\begin{equation}
DS_{V}=dA_{VU}S_{U}+A_{VU}\omega _{U}S_{U}=\omega _{V}S_{V}  \tag{3}
\end{equation}%
where $DA_{VU}=dA_{VU},$ because $A_{VU}$ is the matrix function
transforming $S_{U}$ into $S_{V}$ between two open neighborhoods U$_{U}$ and
U$_{V}$ (U$_{U}$ $\cap $ U$_{V}$ $\neq 0$) on M$^{4}$.

Substituting $A_{VU}^{-1}S_{V}=S_{U}$\textbf{\ }into Eq.(3), we obtain the
relation transforming connection $\omega _{U}$ into connection $\omega _{V}$
as follows \cite{che}

\begin{equation}
\omega _{V}=dA_{VU}A_{VU}^{-1}+A_{VU}\omega _{U}A_{VU}^{-1}  \tag{4}
\end{equation}%
For gauge field (\textsl{\ in principal bundle} ), it is just gauge
transformation between gauge fields $\omega _{U}$ and $\omega _{V}$ in two
open neighborhoods U$_{U}$ and U$_{V}$ (U$_{U}$ $\cap $ U$_{V}$ $\neq 0$) in
gauge field theory; for connection (\textsl{\ in vector bundle }), it is the
transferring relation between connections $\omega _{U}$ and $\omega _{V}$ in
two open neighborhoods U$_{U}$ and U$_{V}$ (U$_{U}$ $\cap $ U$_{V}$ $\neq 0$%
) in curved spacetime, i.e., Eq.(4) is just the unified transformation
expression of gauge field and connection in two open neighborhoods U$_{U}$
and U$_{V}$, thus we can generally call the unified transformation as
general gauge transformation.

Using Eq.(4), we have $\omega _{V}A_{VU}-A_{VU}\omega _{U}=dA_{VU},$ and
further acting the differential connection operator D on Eq.(3), we obtain
not only the unification expression of curvature tensor 2-form (\textsl{\ in
tangent vector bundle} ) and field strength tensor 2-form (\textsl{\ in
principal bundle} ) but also the unification expression of general
coordinate transformation of curvature tensor 2-form (\textsl{\ in tangent
vector bundle} ) and gauge transformation of field strength tensor 2-form (
\textsl{in principal bundle} ) as follows
\begin{equation}
\Omega _{V}=A_{VU}\Omega _{U}A_{VU}^{-1}=d\omega _{V}-\omega _{V}\Lambda
\omega _{V}  \tag{5}
\end{equation}%
In fact, Eqs.(4 \& 5) show that gravity is also a kind of gauge theory.

Multiplying Eq.(5)'s component quantity expression of the curvature tensor
2-form ( \textsl{in tangent vector bundle} )

\begin{equation}
\frac{1}{2}\Omega _{aij}^{\text{ }b}dx^{i}\Lambda dx^{j}=\frac{1}{2}%
A_{a}^{b^{^{\prime }}}\Omega _{b^{^{\prime }}i^{^{\prime }}j^{^{\prime }}}^{%
\text{ }c^{\prime }}A_{\text{ }c^{\prime }}^{-1b}dx^{i^{^{\prime }}}\Lambda
dx^{j^{^{\prime }}}  \tag{6}
\end{equation}%
with $g^{ak}\varepsilon _{bklm}dx^{l}\Lambda dx^{m}$ from right hand side,
in which the a, b, c,$\cdots $, i, j, k, $\cdots $ and a$^{^{\prime }}$, b$%
^{^{\prime }},$ c$^{^{\prime }},\cdots $ i$^{^{\prime }}$, j$^{^{\prime }}$,
k$^{^{\prime }},\cdots $ are in U$_{V}$ and U$_{U}$ respectively, defining $%
\varepsilon _{_{1234}}=(-g)^{\frac{1}{2}}$ \ \cite{hou} \textbf{\&} $%
x^{4}=ict$ and using $\varepsilon _{i_{1}i_{2}i_{3}i_{4}}\varepsilon
^{j_{1}j_{2}j_{3}j_{4}}=sgn(g)\delta
_{i_{1}i_{2}i_{3}i_{4}}^{j_{1}j_{2}j_{3}j_{4}}$ \cite{hou}, we prove

\begin{equation}
\Omega _{ab}^{\text{ }ba}(-g)^{\frac{1}{2}}dx^{1}\Lambda dx^{2}\Lambda
dx^{3}\Lambda dx^{4}=\Omega _{a^{^{\prime }}b^{^{\prime }}}^{\text{ }%
b^{^{\prime }}a^{^{\prime }}}(-g^{\prime })^{\frac{1}{2}}dx^{\prime \text{ }%
1}\Lambda dx^{\prime \text{ }2}\Lambda dx^{\prime \text{ }3}\Lambda
dx^{\prime \text{ }4}  \tag{7}
\end{equation}%
Namely, Eq.(7) is invariant or doesn't depend on coordinates of different U$%
_{V}$ and U$_{U}$.

Adding integral sign to Eq.(7) and, as usual, defining $\Omega _{ab}^{\text{
}ba}=g^{bc}R_{cab}^{a}=R_{V},\Omega _{a^{^{\prime }}b^{^{\prime }}}^{\text{ }%
b^{^{\prime }}a^{^{\prime }}}=R_{U}$ and $d\tau =dx^{1}\Lambda dx^{2}\Lambda
dx^{3}\Lambda dx^{4},$ we obtain the invariant gravitational action \cite%
{car}

\begin{equation}
A=\int_{U_{U}\cap U_{V}}R_{U}(-g)_{_{U}}^{\frac{1}{2}}d\tau
_{U}=\int_{U_{U}\cap U_{V}}R_{V}(-g)_{_{V}}^{\frac{1}{2}}d\tau _{V}  \tag{8}
\end{equation}%
Thus, Eq.(8) keeps invariant in the whole spacetime (manifold M$^{4}=$ $%
\underset{V}{\cup }U_{V}),$ because the two open neighborhoods U$_{U}$ and U$%
_{V}$ are two arbitrary open overlapped neighborhoods in the global
spacetime (\textsl{\ manifold in the vector bundle} ). The above condition
(8) is just the condition that Eq.(8) may be taken as invariant
gravitational action in the whole spacetime, because the physical
consistence demands that Eq.(8) must take the invariant formulation in every
open neighborhood of the whole spacetime, which satisfies just the general
gauge invariant principle of gravitational gauge fields. Then geometry and
physics meanings of gauge invariant property are directly, whole and
seriously revealed.

A Noether invariant quantity is obtained from Eq.(5) in terms of 2-form
field strength tensor ( \textsl{in principal bundle} ) as follows $tr(\Omega
_{V}\Lambda \Omega _{V})=tr(\Omega _{U}\Lambda \Omega _{U}),$which satisfy
just the general gauge invariant principle \cite{wf} of general gauge
fields. Then geometry and physics meanings of gauge invariant property of
general gauge fields are directly, whole and seriously revealed.

Adding integral sign to $tr(\Omega _{V}\Lambda \Omega _{V})$ and using $%
tr(T^{a}T^{b})=-2\delta ^{ab}$, we have the invariant action with dual gauge
field strength tensor in the global curved spacetime as follows \cite{car}

\begin{equation}
A=\int \frac 1{2\pi ^2\kappa }tr(\Omega \Lambda \Omega )=-\int \frac 1{2\pi
^2\kappa }\Omega _{ij}^{\text{ }a}\overset{\symbol{126}}{\Omega }%
^{aij}(-g)^{\frac 12}d\tau  \tag{9}
\end{equation}
Thus the Lagrangian density is $\mathcal{L}=-\frac{(-g)^{\frac 12}}{2\pi
^2\kappa }\Omega _{ij}^{\text{ }a}\overset{\symbol{126}}{\Omega }^{aij}=-%
\frac{(-g)^{\frac 12}}{4q^2}\Omega _{ij}^{\text{ }a}\overset{\symbol{126}}{%
\Omega }^{aij}($ $\overset{\symbol{126}}{\Omega }^{aij}=\frac 12\varepsilon
^{ijkl}\Omega _{kl}^{\text{ }a}$ is dual gauge field strength tensor$,$
taking $\kappa =2q^2/\pi ^2$ $)$ with dual gauge field strength tensor
existing in the global spacetime \cite{NS,car}. As Eq.(9)'s natural
generalization, for higher dimensions, we may naturally deduce $A=\int \frac
1{2\pi ^2\kappa }tr(\Omega \Lambda \Omega \Lambda ...\Lambda \Omega ).$

In terms of modern differential geometry, we can define the inner product $%
<\partial _{\mu },\partial _{\nu }>=<e_{\mu },e_{\nu }>$ $=g_{\mu \nu }$ of
natural (or called intrinsical ) tangent basic vectors and the inner product
$<dx^{\mu },dx^{\nu }>=<e^{\mu },e^{\nu }>$ $=g^{\mu \nu }$ of natural (or
called intrinsical ) cotangent basic vectors$,$ they are consistent, see
Appendix A, then we obtain an important invariant $<\omega _{\mu }dx^{\mu
},\omega _{\nu }dx^{\nu }>=\omega _{\mu }\omega _{\nu }g^{\mu \nu }=\omega
_{\mu }\omega ^{\mu }$ of inner product of two 1-forms (e.g., $\omega
_{V}=\omega _{V\mu }dx_{V}^{\mu })$, i.e., cotangent vectors, under
coordinate transformation. Similar to deducing the important invariant $%
\omega _{\mu }\omega ^{\mu }$ of inner product of two 1-forms, see Appendix
B we finally achieve the invariant action of general gauge fields in the
global curved spacetime as follows

\begin{equation}
A=\int \frac 1{2\pi ^2\kappa }tr<<\Omega ,\Omega >>(-g)^{\frac 12}d\tau
=\int \frac 1{4\pi ^2\kappa }\Omega _{ij}^{\text{ }a}\Omega
^{aij}(-g)^{\frac 12}d\tau  \tag{10}
\end{equation}
Eq.(10) satisfies just the general gauge invariant principle in the global
spacetime.

When $S_{Ui}$ $(i=1,2,3,\cdots ,k;$\textbf{\ }$U$\textbf{\ }is subscript of
the open neighborhood U$_U$\textbf{) }are k linearly independent basic
vectors of the vector space ( \textsl{of vector bundle\ or\ associate vector
bundle }$E(M^4,F,\pi _{_V},G,P(M^4,G,\pi ))$), or equivalently the
orthonormal basic vectors of the Group G's representation vector space F,
Therefore, any vector field S may be expressed as

\begin{equation}
S=S_U\Psi _U  \tag{11}
\end{equation}
where $\Psi _U$ is complex column matrix function of the corresponding
components. It follows from Eq.(2) that

\begin{equation}
D^{^{\prime }}S_{Ui}=\omega _{Uij}^{^{\prime }}S_j=-S_j\omega
_{Uji}^{^{\prime }}  \tag{12}
\end{equation}
in which we have remarked

\begin{equation}
D^{^{\prime }}=\gamma ^\mu D_\mu ,\text{ }d^{^{\prime }}=\gamma ^\mu
\partial _\mu ,\text{ }\omega _{Uij}^{^{\prime }}=\gamma ^\mu \omega _{U\mu
ij}  \tag{13}
\end{equation}
where $\gamma ^\mu $ is $\gamma $ matrix with Lorentz superscript $\mu $ in
Eq.(13), and $dx^{^{\prime }\mu }=a_{\text{ }\nu }^\mu dx^\nu $ $(\mu ,\nu
=0,1,2,3)$ have the same transformation law as $\gamma ^{^{\prime }\mu }=a_{%
\text{ }\nu }^\mu $ $\gamma ^\nu $ \cite{lur}, $a_{\text{ }\nu }^\mu $ is
Lorentz group matrix element. Because the $i,j,..$ and $\mu ,\nu ,...$
belong to the different freedoms' groups, thus, their matrixes corresponding
to different groups are commutable. From Eqs.(2), (11) and (12) we have

\begin{equation}
D^{^{\prime }}S=\omega _U^{^{\prime }}S_U\Psi _U+S_Ud^{^{\prime }}\Psi
_U=S_U(d^{^{\prime }}\Psi _U-\omega _U^{^{\prime }}\Psi _U)  \tag{14}
\end{equation}

Using the orthonormal relation (we can choose the orthonormal basic vectors
of the vector space)

\begin{equation}
(\overset{-}{S}_{Ui},S_{Uj})=\delta _{ij}  \tag{15}
\end{equation}
we obtain

\begin{equation}
(\overline{S},D^{^{\prime }}S)=(\overline{\Psi }_U\overline{S}%
_U,S_U(d^{^{\prime }}\Psi _U-\omega _U^{^{\prime }}\Psi _U))=\overline{\Psi }%
_U(d^{^{\prime }}\Psi _U-\omega _U^{^{\prime }}\Psi _U)  \tag{16}
\end{equation}

Now consider the expression of Eq.(16) in an open neighborhood U$_{W}$. It
follows from $\Psi _{U}=A_{UW}\Psi _{W}$ that

\begin{equation}
d^{\prime }\Psi _U=d^{\prime }A_{UW}\Psi _W+A_{UW}\text{ }d^{\prime }\Psi _W
\tag{17}
\end{equation}

Using Eqs.(4), (16), (17) and $\Psi _{U}=A_{UW}\Psi _{W}$ ( $A_{UW}$ is
transformation matrix ), we have

\begin{equation*}
\overline{\Psi }_U(d^{^{\prime }}\Psi _U-\omega _U^{^{\prime }}\Psi _U)=%
\overline{\Psi }_WA_{UW}^{+}[d^{\prime }A_{UW}\Psi _W+A_{UW}d^{^{\prime
}}\Psi _W
\end{equation*}

\begin{equation}
-(d^{^{\prime }}A_{UW}A_{UW}^{-1}+A_{UW}\omega _{W}^{^{\prime
}}A_{UW}^{-1})A_{UW}\Psi _{W}]=\overline{\Psi }_{W}(d^{^{\prime }}\Psi
_{W}-\omega _{W}^{^{\prime }}\Psi _{W})  \tag{18}
\end{equation}%
where we have used $A_{UW}^{+}A_{UW}=I$. Namely Eq.(18) is the invariant
quantity in the whole spacetime (\textsl{manifold in the bundle}), and the
physical consistence demands that Eq.(18) must take the invariant
formulation in the every open neighborhood of the whole spacetime, which
satisfies just the general gauge invariant principle of the gauge fields
interacting with Fermi fields. Then geometry and physics meanings of gauge
invariant property of gauge fields interacting with Fermion fields are
directly, whole and seriously revealed.

Inserting Eq.(13) into Eq.(18), we obtain the general topological invariant
Lagrangians of matter field $\Psi $ interacting with gauge field $\omega
_{\mu }$, which keeps effective in the whole spacetime as follows

\begin{equation}
\mathcal{L}_\Psi =\overline{\Psi }\gamma ^\mu (\partial _\mu \Psi -\omega
_\mu \Psi )  \tag{19}
\end{equation}

Now we consider another invariant.

Similar to the discussion of Eqs.(14 \& 18), taking $D=dx^{\mu }D_{\mu },$ $%
d=dx^{\mu }\partial _{\mu },$ $\omega _{Uij}=dx^{\mu }\omega _{U\mu ij}$,
replacing spinor field $\Psi _{U}$ with scalar field $\varphi $ in Eq.(14),
it follows that $DS=S_{U}(d\Psi _{U}-\omega _{U}\Psi _{U}),$ further taking
Hermite conjugation of $DS$, and making an inner product, relevant to $(%
\overset{-}{S}_{Ui},S_{Uj})=\delta _{ij}$ and $<dx^{\mu },dx^{\nu }>$ , of
both $DS$ and its conjugation, we get

\begin{equation*}
<(\overline{DS},DS)>=<(\overline{\varphi }_{U}(\overset{\overline{\leftarrow
}}{d}-\overline{\omega }_{U})\overline{S}_{U},S_{U}(d\varphi _{U}-\omega
_{U}\varphi _{U}))>
\end{equation*}%
\begin{equation}
=<\overline{\varphi }_{U}(\overset{\overline{\leftarrow }}{\partial }_{\mu
}-\omega _{\mu U})dx^{\mu },dx^{\nu }(\partial _{\nu }\varphi _{U}-\omega
_{\nu U}\varphi _{U})>  \tag{20}
\end{equation}

Analogous to the discussion of Eq.(18), it is easy to prove that Eq.(20) is
the invariant quantity existing in the whole spacetime .

In terms of modern differential geometry, we again use the inner product $%
<dx^{\mu },dx^{\nu }>=<e^{\mu },e^{\nu }>$ $=g^{\mu \nu }$ of natural
cotangent basic vectors$,$ then we achieve an important invariant Lagrangian
of scalar fields interacting with gauge fields

\begin{equation*}
\mathcal{L}_{\varphi }=<(\overline{DS},DS)>=\overline{\varphi }(\overset{%
\overline{\leftarrow }}{\partial }_{\mu }-\overline{\omega }_{\mu })g^{\mu
\nu }(\partial _{\nu }-\omega _{\nu })\varphi
\end{equation*}%
\begin{equation}
=\overline{\varphi }(\overset{\overline{\leftarrow }}{\partial }_{\mu }-%
\overline{\omega }_{\mu })(\partial ^{\mu }-\omega ^{\mu })\varphi  \tag{21}
\end{equation}%
Thus, Eq.(21) satisfies the physical invariant consistent demand in the
every open neighborhood of the global spacetime. That is, Eq.(21) satisfies
just the general gauge invariant principle of the gauge fields interacting
with scalar fields. Then geometry and physics meanings of gauge invariant
property of gauge fields interacting with Boson fields are directly, whole
and seriously revealed.

On the other hand, due to

\begin{equation}
(\overline{S},S)=(\overline{\varphi }_{U}\overline{S}_{U},S_{U}\varphi _{U})=%
\overline{\varphi }_{U}\varphi _{U}  \tag{22}
\end{equation}%
and using $\varphi _{U}=A_{UW}\varphi _{W},$ it is easy to prove that
Eq.(22) is invariant quantity existing in the whole spacetime. Thus we may
multiply Eq.(22) with mass parameter $m^{n}$ ( $n=2,$ taking $\varphi $ as
boson function$;$ $n=1,$ taking $\varphi $ as fermion function ) to
construct the mass part of the Lagrangian of a system.

\begin{equation}
\mathcal{L}_{m\varphi }=m^n\overline{\varphi }\varphi ,  \tag{23}
\end{equation}

About potential functional of $(\overline{\varphi }_U,\varphi _U)$ due to
the invariance of $(\overline{\varphi }_U,\varphi _U)$ under the general
gauge transformation, it must satisfy

\begin{equation}
V_U(\overline{\varphi }_U,\varphi _U)=V_W(\overline{\varphi }_W,\varphi _W)
\tag{24}
\end{equation}
\textbf{\ }which is the invariant condition in the global spacetime, i.e.,
Eq.(24) satisfies just the general gauge invariant principle of potential
energy. For example, Eq.(22)'s arbitrary combinations may satisfy condition
(24)\textsl{.}\textbf{,} i.e., we may generally take Eq.(22) as the variable
to construct scalar potential functional. e. g., the potential functional V
of SU(2) complex scalar fields
\begin{equation}
V(\overline{\varphi }\varphi )=\lambda ^2(\overline{\varphi }\varphi -\mu
^2)^2.  \tag{25}
\end{equation}

Using Eqs.(8), (10), (19), (21). (23) and (24), we achieve gravitational
action, all general actions of matter fields, spinor fields and scalar
fields interacting with gauge fields in the curved spacetime as follows

\begin{equation}
A=\int_{M^4}(\alpha R+\mathcal{L}_m)(-g)^{\frac 12}d\tau  \tag{26}
\end{equation}

\begin{equation}
A=\int_{M^4}[\alpha R+\overline{\Psi }\gamma ^\mu (\partial _\mu \Psi
-\omega _\mu \Psi )+m\overline{\Psi }\Psi +U(\overline{\Psi },\Psi )-\frac
1{4q_\lambda ^2}F_{\mu \nu }^{\text{ }\lambda a}F^{\lambda a\mu \nu
}](-g)^{\frac 12}d\tau  \tag{27}
\end{equation}

\begin{equation}
A=\int_{M^{4}}[\alpha R+\overline{\varphi }(\overset{\overline{\leftarrow }}{%
\partial ^{\mu }}-\overline{\omega }^{\mu })(\partial _{\mu }-\omega _{\mu
})\varphi +m^{2}\overline{\varphi }\varphi +V(\overline{\varphi },\varphi )-%
\frac{1}{4q_{\lambda }^{2}}F_{\mu \nu }^{\text{ }\lambda a}F^{\lambda a\mu
\nu }](-g)^{\frac{1}{2}}d\tau  \tag{28}
\end{equation}%
where $\alpha $ is a parameter, $\mathcal{L}_{m}$ is the Lagrangian of
general matter fields, $\overline{\omega }^{\mu }=g^{\mu \nu }\overline{%
\omega }_{\nu }$ and $\gamma ^{\mu }=g^{\mu \nu }\gamma _{\nu }$\textbf{. (}%
\textsl{In the case that the fibre G of principal bundle is a semisimple
group of a general form,} ) the Lagrangian contains r arbitrary constants $%
q_{_{\lambda }}$ , $\lambda =1,2,...,r$, in which r is the number of
invariant simple factors \cite{car}.

The fundamental physics interactions, e.g., strong, weak, electromagnetic
and gravitational interactions, can be described by Eqs.(26-27), these
actions are unifiedly deduced by QCP, i.e., Eq.(1), thus the new unification
theory of fundamental physics interactions is given in terms of modern
differential geometry. The more concrete examples are that the known grand
unification theories may be SU(5), S(10) or E$_{6}$ gauge theories, and it
is very easy that their relative more fermion mass terms etc can similarly
be deduced by using the expression which is relevant to spinor's general
gauge invariant quantities.

\section{The Unified Description of Fundamental Physics Interactions and
Noether Theorem}

In the expression (1) of QCP, which has deduced all the physical fundamental
interactions in this paper, for general field variables $X(x)=$ $\{\Psi
(x),\varphi (x),\omega _{\mu }(x),g_{\mu \nu }(x),...,\}$ above, when S is
the actions (26-28), C is unit element and D is infinitesimal transformation
operator of continuous Lie group \cite{htt,dju,liz}

\begin{equation}
x^\mu \rightarrow x^{\prime }\text{ }^\mu \doteq x^\mu +\Delta x^\mu =x^\mu
+\varepsilon _\sigma \tau ^{\mu \sigma }\text{ }(x,X,X\text{,}_\mu )
\tag{29}
\end{equation}
\begin{equation}
X^a(x)\rightarrow X^{\prime \text{ }a}(x^{\prime })\doteq X^a(x)+\Delta
X^a(x)=X^a(x)+\varepsilon _\sigma \xi ^{a\sigma }(x,X,X\text{,}_\mu )
\tag{30}
\end{equation}
in which $\varepsilon _\sigma $ $(\sigma =1,2,\cdots ,m)$ are infinitesimal
parameters of Lie group \textsl{D}$_m$ in Eq.(1). Thus, we get the unified
expression of their variational principles, as follows

\begin{equation}
\Delta A=DA-A=A^{\prime }-A=0  \tag{31}
\end{equation}%
Not losing generality, under the transformations of Eqs.(29) and (30), we
can take ad-hoc

\begin{equation}
L^{\prime }(x^{\prime },X^{\prime }(x^{\prime }),X^{\prime }\text{,}_{\mu
}(x^{\prime }),X^{\prime }\text{,}_{\mu \nu }(x^{\prime }))=L(x^{\prime
},X^{\prime }(x^{\prime }),X^{\prime }\text{,}_{\mu }(x^{\prime }),X^{\prime
}\text{,}_{\mu \nu }(x^{\prime }))+\partial _{\mu }\Omega ^{\mu }  \tag{32}
\end{equation}%
where $\partial _{\mu }\Omega ^{\mu }=\varepsilon _{\sigma }\partial _{\mu
}\Omega ^{\sigma \mu }$ and the Ricci Scalar R contains $g_{\alpha \beta }$,$%
_{\mu \nu }$, and $\Omega$ may be rapidly decreasing to fit the usual
physics experiments.

Using the unified expression of the variational principles, we have%
\begin{equation*}
\Delta A=\int_{M^{4}}\{[\frac{\partial L}{\partial X^{a}}-\partial _{\mu }%
\frac{\partial L}{\partial X^{a}\text{,}_{\mu }}+\partial _{\mu }\partial
_{\nu }\frac{\partial L}{\partial X^{a}\text{,}_{\mu \nu }}]\delta X^{a}+
\end{equation*}

\begin{equation}
\partial _{\mu }[L\Delta x^{\mu }+(\frac{\partial L}{\partial X^{a}\text{,}%
_{\mu }}-\partial _{\nu }\frac{\partial L}{\partial X^{a}\text{,}_{\mu \nu }}%
)\delta X^{a}+\frac{\partial L}{\partial X^{a}\text{,}_{\mu \nu }}\delta
X^{a}\text{,}_{\nu }+\Omega ^{\mu }]\}d^{4}x  \tag{33}
\end{equation}%
in which $\delta X^{a}=\Delta X^{a}-X^{a}$,$_{\nu }\Delta x^{\nu }$. \textbf{%
U}sing the above discussions, we can obtain their two-order general
Euler-Lagrange Equations and corresponding Noether conservation currents as
follows

\begin{equation}
\frac{\partial L}{\partial X^a}-\partial _\mu \frac{\partial L}{\partial X^a%
\text{,}_\mu }+\partial _\mu \partial _\nu \frac{\partial L}{\partial X^a%
\text{,}_{\mu \nu }}=0  \tag{34}
\end{equation}
\begin{equation}
\partial _\mu J^{\mu \sigma }=0  \tag{35}
\end{equation}
\begin{equation}
J^{\mu \sigma }=L\tau ^{\mu \sigma }+(\frac{\partial L}{\partial X^a\text{,}%
_\mu }-\partial _\nu \frac{\partial L}{\partial X^a\text{,}_{\mu \nu }})(\xi
^{a\sigma }-X^a,_\alpha \tau ^{\alpha \sigma })+\frac{\partial L}{\partial
X^a\text{,}_{\mu \nu }}\partial _\nu (\xi ^{a\sigma }-X^a,_\alpha \tau
^{\alpha \sigma })+\Omega ^{\mu \sigma }  \tag{36}
\end{equation}
Therefore, Noether theorem of the general physical system is deduced by QCP%
\textbf{. }

In all, using QCP, i.e., Eq.(1), we derive not only all the fundamental
physics Lagrangians and their variational principles and further Noether
theorem\textbf{,} but also their two-order general Euler-Lagrange Equations
and corresponding Noether conservation currents.\ Using the above studies,
variation laws of the different physical systems are naturally determined.
Therefore, it is essential to deduce effectively unified expressions of
elementary physical laws by QCP.

\section{\protect\bigskip Applications and New Physics}

We can use the geometric unification theory to study different physical
systems. For example:

(1) As Eq.(24)'s natural generalization, when the invariant scalar product $%
\overline{\varphi }\varphi $ is extended by the invariant scalar curvature $%
R,$ we may naturally deduce $f(R,\overline{\varphi }\varphi )$ gravity,
i.e., $A=\int f(R,\overline{\varphi }\varphi )(-g)^{\frac{1}{2}}d\tau ,$
because$(-g)^{\frac{1}{2}}d\tau $ is invariant volume element. Further
adding the first term $\alpha R$ of Eq.(28), we get $A_{1}=\int F(R,%
\overline{\varphi }\varphi )(-g)^{\frac{1}{2}}d\tau =$ $\int [f(R,\overline{%
\varphi }\varphi )+\alpha R](-g)^{\frac{1}{2}}d\tau $ \cite{rev}.

(2) As Eq.(24)'s direct application, when the invariant scalar product $%
\overline{\varphi }\varphi $ is taken as the invariant scalar product $%
H^{+}H $ of Higgs field H's \{5\} multiplets, we may naturally deduce $V(H)$
$=\frac{\mu _{0}^{2}}{2}H^{+}H+\frac{\lambda }{4}(H^{+}H)^{2}$ ($\mu _{0}$
and $\lambda $ are parameters), which is just the potential of Higgs field
H's \{5\} multiplets for grand unification theory \cite{rev1}; for Higgs
field $\phi $'s \{24\} multiplets that are expressed by a $5\times 5$
matrix, the matrix satisfies Eq.(5)'s general gauge invariant property,
i.e., $\phi _{V}=A_{VU}\phi _{U}A_{VU}^{-1}$ or $tr\phi _{V}=tr\phi _{U}$
and more generally it follows that $tr\phi _{V}^{n}=tr\phi _{U}^{n}$ , and
as Eq.(24)'s natural generalization, we have $V(\phi )$ $=$ $-\frac{1}{2}\mu
^{2}tr\phi ^{2}+\frac{a}{4}(tr\phi ^{2})^{2}+\frac{b}{2}tr\phi ^{4}$ ( $\mu
, $ $a$ and $b$ are parameters ); for mixed invariants of H's \{5\} and $%
\phi $'s \{24\} multiplets, similarly, people can deduce $V(H,\phi
)=cH^{+}H(tr\phi ^{2})+dH^{+}\phi ^{2}H$ ( $c$ and $d$ are parameters ),
thus we totally deduce $V_{t}(H,\phi )=V(H)+V(\phi )+V(H,\phi ),$ which is
the total Higgs potential of grand unification theory and the same as that
of Ref. \cite{rev1}.

(3) Generalizing $R$ as an invariant functional $f(R)$ and using Eq.(27)
deduced by QCP, people can deduce Fermion field Lagrangian interacting with
Non-Abelian gauge field and general $f(R)$ gravitational field as follows
\begin{equation}
A=\int_{M^{4}}[\alpha f(R)+F_{\Psi }(\overline{\Psi }\gamma ^{\mu }(\partial
_{\mu }\Psi -\omega _{\mu }\Psi ),\overline{\Psi }\Psi )+U(\overline{\Psi }%
,\Psi )+F_{A}(F_{\mu \nu }^{\text{ }\lambda a}F^{\lambda a\mu \nu })](-g)^{%
\frac{1}{2}}d\tau .  \tag{37}
\end{equation}

Eq.(37) is just a general generalization of Eq.(27), and further utilizing
the deduced Eqs.(34) and (35) by QCP, people can naturally gives the new
unification theory of the fundamental Non-Abelian gauge field, Fermion field
and general $f(R)$ gravitational field interactions and Noether theorem by
using QCP, and can concretely derive their Euler-Lagrange Equations and
corresponding Noether conservation currents as done in Sect. 3. Thus the
relative books and articles may be renewedly and systematically rewritten,
which will help people to understand and express the fundamental Non-Abelian
gauge field, Fermion field and general $f(R)$ gravitational field
interactions simpler and with clear quantitative causal physical meanings.

(4) Generalizing $R$ as an invariant functional $f(R)$ and using Eq.(28)
deduced by QCP, people can deduce Boson field Lagrangian interacting with
Non-Abelian gauge field and general $f(R)$ gravitational field as follows
\begin{equation}
A=\int_{M^{4}}[\alpha f(R)+F_{\varphi }(\overline{\varphi }(\overset{%
\overline{\leftarrow }}{\partial ^{\mu }}-\overline{\omega }^{\mu
})(\partial _{\mu }-\omega _{\mu })\varphi ,\overline{\varphi }\varphi )+V(%
\overline{\varphi },\varphi )+F_{A}(F_{\mu \nu }^{\text{ }\lambda
a}F^{\lambda a\mu \nu })](-g)^{\frac{1}{2}}d\tau .  \tag{38}
\end{equation}

Eq.(38) is just a general generalization of Eq.(28), and further utilizing
the deduced Eqs(34) and (35) by using QCP, people can naturally gives new
unification theory of the fundamental Non-Abelian gauge field and general $%
f(R)$ gravitational field interactions and Noether theorem by using QCP, and
can concretely derive their Euler-Lagrange Equations and corresponding
Noether conservation currents. Thus the relevant books and articles may be
renewedly and systematically rewritten, which will help people to understand
and express the fundamental Non-Abelian gauge field, Boson field and general
$f(R)$ gravitational interactions simpler and with clear quantitative causal
physical meanings. When $f(R)=0,$ we naturally give electric and color
superconduction theories corresponding Abelian $-\frac{1}{4q^{2}}F_{\mu \nu
}^{\text{ }}F^{\mu \nu }$ and non-Abelian $-\frac{1}{4q_{\lambda }^{2}}%
F_{\mu \nu }^{\text{ }\lambda a}F^{\lambda a\mu \nu }$ in Eq.(38)$,$
respectively \cite{rev1}, thus the relevant books and articles may be
renewedly and systematically rewritten, which will help people to understand
and express the fundamental Non-Abelian gauge field and Boson field
interactions simpler and with clear quantitative causal physical meanings.

(5) Using QCP, people can more generally deduce%
\begin{equation*}
A=\int_{M^{4}}[\alpha f(R)+\overline{\Psi }\gamma ^{\mu }(\partial _{\mu
}\Psi -\omega _{\mu }\Psi )+m_{\Psi }\overline{\Psi }\Psi +\overline{\varphi
}(\overset{\overline{\leftarrow }}{\partial ^{\mu }}-\overline{\omega }^{\mu
})(\partial _{\mu }-\omega _{\mu })\varphi
\end{equation*}%
\begin{equation}
+m_{\varphi }^{2}\overline{\varphi }\varphi +M(U(\overline{\Psi },\Psi ),V(%
\overline{\varphi },\varphi ))+V_{t}(H,\phi )-\frac{1}{4q_{\lambda }^{2}}%
F_{\mu \nu }^{\text{ }\lambda a}F^{\lambda a\mu \nu }](-g)^{\frac{1}{2}%
}d\tau .  \tag{39}
\end{equation}

Eq.(39) is just a general generalization of Eqs.(37) and (38), and further
utilizing the deduced Eqs(34) and (35) by QCP, people can naturally gives
the new unification description theory of the fundamental Non-Abelian gauge
field, Fermion field, Boson field, general $f(R)$ gravitational field
interactions and Noether theorem by using QCP, and can concretely derive
their Euler-Lagrange Equations and corresponding Noether conservation
currents. Thus the relevant books and articles may be renewedly and
systematically rewritten, which will make people understand and express the
fundamental Non-Abelian gauge field, Fermion field, Boson field, and general
$f(R)$ gravitational field interactions simpler and with clear quantitative
causal physical meanings. Specially, Eq.(39) is just the action of the
theory of unifiedly describing all known physical fundamental interactions,
further further utilizing the deduced Eqs(34) and (35) by QCP, consequently
the new unification theory of the fundamental physics interactions and
Noether theorem is given by using QCP for the first time.

When considering nonlinear dynamic interaction action of their fields in
Eq.(39) in current particle physics standard model, we deduce

\begin{eqnarray*}
A &=&\int_{M^{4}}[\alpha f(R)+F_{\Psi }(\overline{\Psi }\gamma ^{\mu
}(\partial _{\mu }\Psi -\omega _{\mu }\Psi ),\overline{\Psi }\Psi ) \\
&&+F_{\varphi }(\overline{\varphi }(\overset{\overline{\leftarrow }}{%
\partial ^{\mu }}-\overline{\omega }^{\mu })(\partial _{\mu }-\omega _{\mu
})\varphi ,\overline{\varphi }\varphi )
\end{eqnarray*}%
\begin{equation}
+M(U(\overline{\Psi },\Psi ),V(\overline{\varphi },\varphi ))+V_{t}(H,\phi
)+F_{A}(F_{\mu \nu }^{\text{ }a}F^{a\mu \nu })](-g)^{\frac{1}{2}}d\tau .
\tag{40}
\end{equation}

When further considering two types of nonlinear dynamic interaction actions
in Eq.(40) in current particle physics standard model, we have

\begin{eqnarray*}
A &=&\int_{M^{4}}[\alpha f(R)+F_{R\Psi }(R,\overline{\Psi }\gamma ^{\mu
}(\partial _{\mu }\Psi -\omega _{\mu }\Psi ),\overline{\Psi }\Psi )+F_{\Psi
}(\overline{\Psi }\gamma ^{\mu }(\partial _{\mu }\Psi -\omega _{\mu }\Psi ),%
\overline{\Psi }\Psi ) \\
&&+F_{R\varphi }(R,\overline{\varphi }(\overset{\overline{\leftarrow }}{%
\partial ^{\mu }}-\overline{\omega }^{\mu })(\partial _{\mu }-\omega _{\mu
})\varphi ,\overline{\varphi }\varphi )+F_{\varphi }(\overline{\varphi }(%
\overset{\overline{\leftarrow }}{\partial ^{\mu }}-\overline{\omega }^{\mu
})(\partial _{\mu }-\omega _{\mu })\varphi ,\overline{\varphi }\varphi )
\end{eqnarray*}%
\begin{eqnarray*}
&&+F_{\Psi \varphi }(\overline{\Psi }\gamma ^{\mu }(\partial _{\mu }\Psi
-\omega _{\mu }\Psi ),\overline{\Psi }\Psi ,\overline{\varphi }(\overset{%
\overline{\leftarrow }}{\partial ^{\mu }}-\overline{\omega }^{\mu
})(\partial _{\mu }-\omega _{\mu })\varphi ,\overline{\varphi }\varphi ) \\
&&+F_{RA}(R,F_{\mu \nu }^{\text{ }a}F^{a\mu \nu })+F_{\Psi A}(\overline{\Psi
}\gamma ^{\mu }(\partial _{\mu }\Psi -\omega _{\mu }\Psi ),\overline{\Psi }%
\Psi ,F_{\mu \nu }^{\text{ }a}F^{a\mu \nu }) \\
&&+F_{\varphi A}(\overline{\varphi }(\overset{\overline{\leftarrow }}{%
\partial ^{\mu }}-\overline{\omega }^{\mu })(\partial _{\mu }-\omega _{\mu
})\varphi ,\overline{\varphi }\varphi ,F_{\mu \nu }^{\text{ }a}F^{a\mu \nu })
\end{eqnarray*}%
\begin{equation}
+M(U(\overline{\Psi },\Psi ),V(\overline{\varphi },\varphi ))+V_{t}(H,\phi
)+F_{A}(F_{\mu \nu }^{\text{ }a}F^{a\mu \nu })](-g)^{\frac{1}{2}}d\tau .
\tag{41}
\end{equation}

Using quantitative causal principle, we not only can give a restatement of
well-known physical theories/Lagrangians in the language of fibre bundles,
but also give the new unified description of the fundamental physics
interactions and Noether theorem in this paper, these are new physics and
cannot be given in the past \cite{rev1}.

Especially, Eqs.(40) and (41) give a lot of new physics and new physical
interactions in current particle physics standard model, which can describe
a lots of high order fundamental interactions and different nonlinear
physical systems in different branches of physics, e.g., high order
fundamental interactions (especially in both early universe and particle
physics standard model), cosmology, strong field physics, condensed matter
physics, nonlinear systems etc.

On the other hand, using the research in this paper, we naturally have the
least coupling action between gravitational curvature tensor and gauge field
strength tensor in current particle physics standard model

\begin{equation*}
A=\int_{M^{4}}[\eta _{0}R-\frac{1}{4q_{\lambda }^{2}}F_{\mu \nu }^{\text{ }%
\lambda a}F^{\lambda a\mu \nu }-\frac{\eta _{1}}{4}Tr(R_{\mu \nu \alpha
\beta }F^{\mu \nu }F^{\alpha \beta })](-g)^{\frac{1}{2}}d\tau
\end{equation*}%
\begin{equation}
=\int_{M^{4}}[\eta _{0}R-\frac{1}{4q_{\lambda }^{2}}F_{\mu \nu }^{\text{ }%
\lambda a}F^{\lambda a\mu \nu }+\frac{\eta _{1}}{2}R_{\mu \nu \alpha \beta
}F^{b\mu \nu }F^{b\alpha \beta }](-g)^{\frac{1}{2}}d\tau .  \tag{42}
\end{equation}

Eq.(42) gives new interaction vertexes, new Feynman figures between
gravitational field and gauge fields with U(1), SU(2), SU(3), SU(5), SO(10)
etc symmetries, respectively, which are new physics and cannot given in the
past \cite{rev1}.

When considering high order fundamental interactions and non-linear
dynamical systems, Eq.(42) can be generally written as

\begin{equation}
A=\int_{M^{4}}[\alpha f(R)+F_{A}(F_{\mu \nu }^{\text{ }a}F^{a\mu \nu
})+F_{R^{\prime }A^{\prime }}(R_{\mu \nu \alpha \beta }F^{b\mu \nu
}F^{b\alpha \beta })](-g)^{\frac{1}{2}}d\tau .  \tag{43}
\end{equation}

Further using Eq.(40), we have

\begin{eqnarray*}
A &=&\int_{M^{4}}[\alpha f(R)+F_{\Psi R^{\prime }A^{\prime }}(\overline{\Psi
}\gamma ^{\mu }(\partial _{\mu }\Psi -\omega _{\mu }\Psi ),\overline{\Psi }%
\Psi ,R_{\mu \nu \alpha \beta }F^{b\mu \nu }F^{b\alpha \beta }) \\
&&+F_{\varphi R^{\prime }A^{\prime }}(\overline{\varphi }(\overset{\overline{%
\leftarrow }}{\partial ^{\mu }}-\overline{\omega }^{\mu })(\partial _{\mu
}-\omega _{\mu })\varphi ,\overline{\varphi }\varphi ,R_{\mu \nu \alpha
\beta }F^{b\mu \nu }F^{b\alpha \beta })
\end{eqnarray*}%
\begin{equation}
+M(U(\overline{\Psi },\Psi ),V(\overline{\varphi },\varphi ))+V_{t}(H,\phi
)+F_{A}(F_{\mu \nu }^{\text{ }a}F^{a\mu \nu })](-g)^{\frac{1}{2}}d\tau .
\tag{44}
\end{equation}

\bigskip Further considering Weyl tensor $W^{\mu \nu \alpha \beta }$ and $%
\overset{\symbol{126}}{F}^{\alpha \beta }=\frac{1}{2}\varepsilon ^{\alpha
\beta \mu \nu }F_{\mu \nu }$, Eq.(43) can be generalized as

\begin{equation}
A=\int_{M^{4}}[\alpha f(R)+F_{1}(F_{\mu \nu }F^{\mu \nu })+F_{2}(F_{\alpha
\beta }\overset{\symbol{126}}{F}^{\alpha \beta })+F_{3}(F_{\mu \nu
}F_{\alpha \beta }W^{\mu \nu \alpha \beta })](-g)^{\frac{1}{2}}dx^{4}.
\tag{45}
\end{equation}

In Eq.(45), when taking special expressions $\alpha f(R)=\frac{1}{2M_{PI}^{2}%
}R$, $F_{1}(F_{\mu \nu }F^{\mu \nu })=-\frac{1}{4}F_{\mu \nu }F^{\mu \nu }+%
\frac{\alpha _{1}}{4M_{PI}^{4}}(F_{\mu \nu }F^{\mu \nu })^{2}$, $%
F_{2}(F_{\alpha \beta }\overset{\symbol{126}}{F}^{\alpha \beta })=\frac{%
\alpha _{2}}{4M_{PI}^{4}}(F_{\alpha \beta }\overset{\symbol{126}}{F}^{\alpha
\beta })^{2}$, $F_{3}(F_{\mu \nu }F_{\alpha \beta }W^{\mu \nu \alpha \beta
})=\frac{\alpha _{3}}{2M_{PI}^{2}}F_{\mu \nu }F_{\alpha \beta }W^{\mu \nu
\alpha \beta }$, we deduce the key formula (2) in the important Ref.\cite%
{weak}. Thus our research is consistent with current research.

All the deduced Lagrangians in this paper in current particle physics
standard model are invariant in all general coordinate transformations (for
tangent bundle) and gauge transformations (for principal bundle), which just
means that the physics laws are not dependent on people's observations.

The general Lagrangians of both high order fundamental interactions and
different nonlinear physical interactions in current particle physics
standard model cannot be given up to now \cite{rev1}. Therefore, this paper,
for the first time, gives all these Lagrangians, thus we may give these
expression's new physics, e.g., relevant to Eqs. (37), (38) and (40-44).

Einstein thought, god would not care about only inertial system, and god
should specially care about the invariable non-inertial system under general
coordinate transformation, because only using the general non-inertial
system people can identify the most general laws of physics, and can deduce
that the general laws of physics do not depend on people's observation, god
should be more capable, so Einstein believes that the general theory of
relativity is correct.

Similarly, the investigations in this paper are the same. Our investigations
also satisfy the invariances of general coordinate transformation and gauge
transformation, and also lead to the fact that physical laws do not depend
on human observation. God should be more capable and would not give up such
a good plan not to choose.

Especially so far, no experiments prove the existence of a higher
dimensional space \cite{rev1}, and this studies of this paper do not need
the extra higher dimensional spaces, we realize the unification of the
fundamental interaction laws of physics, this theory is the
\textquotedblleft from down to up\textquotedblright\ beautiful
phenomenological theory of the very economical and very realistic
symmetries, so the research of this paper is of important theoretical and
practical values, and many new physical predictions are presented.

For example, this paper discovers the new Lagrangians in the equations (37),
(38) and (40-44), their corresponding Euler-Lagrange equations by Eq.(34)
and the new physical interaction law expressed by the equations and so on.

Because of length limitation of the paper, the other Lagrangians in physics
and the corresponding new physics may be analogously derived, their detailed
applications can be given easily by means of the research of this paper and
which will be written in our following papers.

This principle not only is not opposed to the usual restrictions of
causality in a quantum field theory, but also the principle is more general,
and can deduce the usual restrictions of causality in a quantum field
theory. Because the usual restrictions of causality in a quantum field
theory are mainly coming from the restrictions of Lorentz transformations,
Lorentz transformations consist of some equations, any equation must satisfy
quantitative causal principle, namely, some changes (cause) of some
quantities in any equation must result in the relative changes (result) of
the other quantities in the equation so that the equation's right side keeps
no-loss-no-gain, i.e., zero, namely, the equation satisfies the quantitative
causal principle, which is very similar to Eq.(2.4)'s casual discussion
(below Eq.(2.4) in Ref.\cite{liao}.

Causality is one of the most fundamental and essential notions in physics
\cite{Einst}. Causal efficacy cannot propagate faster than light, which
comes from the constraints of Lorentz transformations of special relativity.
Thus, in space-like spacetime, any two physical operators in a general
quantum field are commutative, because they cannot have any relation by
Lorentz transformations.

For symmetric principle \cite{symm}, people currently know that it is the
fundamental principle of physics, which may deduce various symmetric groups,
four fundamental interactions of physics and so on \cite{wei,liz,car,perski}%
. Consequently, symmetry is playing the key role in physics.

Actually, Eq.(1) is just a special situation of symmetric principle, namely,
the system is invariant under some operations, i.e., symmetric
transformations that keep quantitative causal relation to be invariant.
Consequently, QCP is a special expression of symmetric principle, which
means that symmetric principle is more general and includes QCP as a subset.
Thus this paper deduces QCP from symmetric principle.

But in the past, QCP and symmetric principle are firmly viewed as two
independent fundamental principles, this paper gives the unification of QCP
and symmetric principle, which will hugely affect the global physics and the
corresponding developments relevant to QCP and symmetric principle in both
physics and science, because causal principle and symmetric principle both
are  the very old, classical, famous and fundamental principles, the two
principles have been hugely affecting the developments of different branches
of physics and science for very long time.

\section{\protect\bigskip Summary and Conclusions}

This paper discovers and naturally gives geometric unification theory of
the grand unification and gravitational interactions, and shows
unification description theory of the grand unifying and gravitational interactions and Noether theorem by utilizing QCP, i.e., Eq.(1). The
invariant quantities covering all the parts of the Lagrangians of the four
fundamental physics interactions are deduced by QCP, all these invariant
quantities satisfy general gauge invariance of the general gauge fields
interacting with Fermion and/or boson fields. These different invariant
quantities are constructed ( \textsl{in vector bundle or principal bundle,
whose base manifolds are a 4-dimensional physical general curved spacetime
manifold M}$^{4}$). Thus, using these invariant quantities, all the
Lagrangians of the four physical fundamental interactions are given and
satisfy the general gauge invariant principle of general gauge fields
interacting with Fermion and/or boson fields, and geometry and physics
meanings of gauge invariant property of the different physical systems are
directly, whole and seriously revealed.

Actually it is in terms of QCP that their variational principles and
corresponding Noether theorem in field theory are derived. Therefore, the
very hard difficulty ( that the four fundamental physics interactions and
variational principle, further Noether theorem, previously regarded as
independent fundamental laws, were not able to be unifiedly proved and
investigated in the past \cite{perski}) is naturally overcome in this paper.
Namely, the geometric unification theory and origins of the four fundamental
physics interactions and the Noether theorems are shown by QCP, furthermore,
their two-order general Euler-Lagrange equations and corresponding Noether
conservation currents are obtained.

In fact, general physics process \cite{hdf,htt,uni} and different physics
processes \cite{haa,hbb,cau,ger} all satisfy QCP with no-loss-no-gain
character, the above investigations satisfy QCP, and are consistent.
Especially, using the theory of this paper, a lot of research works about
different branches of physics etc can be anew done and expressed simpler
with clear quantitative causal physical meanings, e.g., people can deduce a
lot of different parts of different Lagrangians, which will naturally give
Lagrangians of different physics systems according to QCP and different
symmetries etc, and then using Eqs.(34-36) people can give all corresponding
concrete physics laws, i.e., find new Euler-Lagrange equations and
conservation laws, and discover new concrete physics laws and so on. Thus
the theory of this paper is useful and will be broad utilized and cited in
different branches of physics and so on due to the key important properties
of the four fundamental interactions for different physics systems.
Therefore, all articles and (text)books relevant to the four fundamental
physics interactions and Noether theorem would be supplemented with the new
conclusions, and this paper will be cited by using the geometric unification
theory in different physical systems, e.g., condensed physics, atomic
physics, molecular physics, quantum optics, nuclear physics, particle
physics, quantum field theories and so on.

Because the standard model of strong, weak and electromagnetic interactions
can be included in a unification interaction model of a big gauge group,
thus people get the grand unified theory, and because the gravitational
interaction is related to the structures of spacetime, and the spacetime is
the stage performing the laws of strong, weak and electromagnetic
interactions, gravitational interaction cannot be included into a
unification interaction model of a larger gauge group. This is because the
grand unification theory of gauge theories of strong, weak and
electromagnetic interactions is based on principal bundle theory in
differential geometry, and gravitational theory is based on the tangent
vector bundle theory in differential geometry \cite{hou}, so people cannot
unify the these four basic interactions in principal bundle theory in
differential geometry theory, so for their unification, people have met
great difficulties.

In order to overcome these difficulties, people have to increase the
dimensional number of spacetime etc to solve these problems, the most
promising success theory is superstring theory, but in return to low
four-dimensional case etc, superstring theory encounters the big
difficulties \cite{bbs}, up to now, there is no way that, in the case of
four dimensions, strong, weak, electromagnetic and gravitational
interactions are unified as one unification description theory.

This paper, for the first time, discovers and gives geometric unification
theory of strong, weak, electromagnetic and gravitational interactions by
investigating the general fiber bundle theory ( in differential geometry
theory ) of unifiedly describing both principal bundle theory and tangent
bundle theory and by studying important laws in physics, e.g., relations
between symmetry and QCP etc. Consequently, the research of this paper is
based on the rigorous scientific bases of mathematics and physics,
therefore, people can essentially and really understand the four basic
interaction laws of physics as well as the relations between them in terms
of the research in this paper, for their further development builds up the
scientific solid foundation.

This paper also discovers the new Lagrangians in Eqs.(37), (38) and (40-44),
their corresponding Euler-Lagrange equations by Eq.(34) and the new physical
interaction law expressed by the equations and so on, i.e., gives the
relevant new physics.

This paper deduces QCP from symmetric principle, and gives the unification
of QCP and symmetric principle, which would hugely affect the physics and
the corresponding developments in both physics and science, because both
causal principle and symmetric principle are the very old and fundamental
principles, and the two principles have been hugely affecting the
developments of different branches of physics and science for very long
time. Thus, this paper opens a door to both study and give new developments
of geometric unification theory of physics laws, and using the new geometric
unification theory, a lot of research works about different branches of
physics can be anew done and expressed simpler with different symmetric
characters.

Acknowledgments

Authors are grateful to Prof. Z. P. Li for useful discussion and comment.

The work is supported by NSF through grants PHY-0805948, DOE USA through
grant DE-SC0007884, and National Natural Science Foundation of China ( No.
11875081 ).

Appendix A

For any natural coordinate $x^\mu $ and local orthogonal coordinate $x^a $
in any 4-dimensional general spacetime, any tangent vector along any curve
line $t$ can be formally written as $X=\frac d{dt}=\frac{dx^\mu }{dt}\frac
\partial {\partial x^\mu }=\frac{dx^a}{dt}\frac \partial {\partial x^a},$
then the inner product of the tangent vector equates

\begin{equation}
<X,X>=\frac{dx^{\mu }}{dt}\frac{dx^{\nu }}{dt}<\frac{\partial }{\partial
x^{\mu }},\frac{\partial }{\partial x^{\nu }}>=\frac{dx^{a}}{dt}\frac{dx^{b}%
}{dt}<\frac{\partial }{\partial x^{a}},\frac{\partial }{\partial x^{b}}>
\tag{A1}
\end{equation}%
where we may define the inner product $<\partial _{\mu },\partial _{\nu
}>=<e_{\mu },e_{\nu }>$ $=g_{\mu \nu }$ ( i.e., natural covariant metric )
of natural tangent basic vectors and the inner product $<\partial
_{a},\partial _{b}>=<e_{a},e_{b}>$ $=\eta _{ab}$ ($\eta _{ab}=-1,$ as $%
a=b=0; $ $\eta _{ab}=1,$ as $a=b=1,2,3;$ $\eta _{ab}=0,$ as $a\neq b,$ i.e.,
orthonormal covariant metric ) of orthonormal tangent basic vectors, so we
can deduce the infinitesimal line element square
\begin{equation}
ds^{2}=<X,X>(dt)^{2}=g_{\mu \nu }dx^{\mu }dx^{\nu }=\eta _{ab}dx^{a}dx^{b}
\tag{A2}
\end{equation}

Similar to $<\partial _{\mu },\partial _{\nu }>=g_{\mu \nu }$ and $<\partial
_{a},\partial _{b}>$ $=\eta _{ab},$ we define the inner product of natural
(or called intrinsical ) cotangent basic vectors as $<dx^{\mu },dx^{\nu
}>=<e^{\mu },e^{\nu }>$ $=g^{\mu \nu }$ ( i.e., natural contravariant metric
) and the inner product of orthonormal cotangent basic vectors as $%
<dx^{a},dx^{b}>=<e^{a},e^{b}>$ $=\eta ^{ab}$ (($\eta ^{ab}=-1,$ as $a=b=0;$ $%
\eta ^{ab}=1,$ as $a=b=1,2,3;$ $\eta ^{ab}=0,$ as $a\neq b,$ i.e.,
orthonormal contravariant metric )$,$ thus, we prove
\begin{eqnarray}
g_{\mu \nu }g^{\mu \alpha } &=&<\frac{\partial }{\partial x^{\mu }},\frac{%
\partial }{\partial x^{\nu }}><dx^{\mu },dx^{\alpha }>  \notag \\
&=&\frac{\partial x^{a}}{\partial x^{\mu }}\frac{\partial x^{b}}{\partial
x^{\nu }}<\frac{\partial }{\partial x^{a}},\frac{\partial }{\partial x^{b}}>%
\frac{dx^{\mu }}{dx^{c}}\frac{dx^{\alpha }}{dx^{d}}<dx^{c},dx^{d}>  \notag
\end{eqnarray}%
\begin{equation}
=\frac{\partial x^{a}}{\partial x^{c}}\frac{\partial x^{b}}{\partial x^{\nu }%
}\eta _{ab}\frac{dx^{\alpha }}{dx^{d}}\eta ^{cd}=\delta _{c}^{a}\frac{%
\partial x^{b}}{\partial x^{\nu }}\eta _{ab}\frac{dx^{\alpha }}{dx^{d}}\eta
^{cd}=\delta _{\nu }^{\alpha }  \tag{A3}
\end{equation}

Furthermore, using Eq.(A2) we deduce $ds^2=g_{\mu \nu }dx^\mu dx^\nu =\eta
_{ab}\frac{\partial x^a}{\partial x^\mu }\frac{\partial x^b}{\partial x^\nu }%
dx^\mu dx^\nu =\eta _{ab}e_\mu ^ae_\nu ^bdx^\mu dx^\nu ,$ i.e., we prove

\begin{equation}
g_{\mu \nu }=\eta _{ab}\frac{\partial x^{a}}{\partial x^{\mu }}\frac{%
\partial x^{b}}{\partial x^{\nu }}=\eta _{ab}e_{\mu }^{a}e_{\nu }^{b}
\tag{A4}
\end{equation}%
where $\frac{\partial x^{a}}{\partial x^{\mu }}=e_{\mu }^{a}$ is vierbein.
On the other hand, we may directly prove Eq.(A4) by using $g_{\mu \nu }=<%
\frac{\partial }{\partial x^{\mu }},\frac{\partial }{\partial x^{\nu }}>=%
\frac{\partial x^{a}}{\partial x^{\mu }}\frac{\partial x^{b}}{\partial
x^{\nu }}<\frac{\partial }{\partial x^{a}},\frac{\partial }{\partial x^{b}}%
>=\eta _{ab}e_{\mu }^{a}e_{\nu }^{b},$ and we may have $g^{\mu \nu
}=<dx^{\mu },dx^{\nu }>=\frac{dx^{\mu }}{dx^{c}}\frac{dx^{\nu }}{dx^{d}}%
<dx^{c},dx^{d}>=e_{c}^{\mu }e_{d}^{\nu }\eta ^{cd}.$ Thus this method is
very convenient.

We still need to prove the consistence of indexes' raising and lowing by
using covariant and contravariant metrics as follows

\begin{equation*}
A^{\mu }=g^{\mu \nu }A_{\nu }=<dx^{\mu },dx^{\nu }>A_{\nu }=\frac{dx^{\mu }}{%
dx^{c}}\frac{dx^{\nu }}{dx^{d}}<dx^{c},dx^{d}>A_{\nu }
\end{equation*}%
\begin{equation}
=e_{c}^{\mu }e_{d}^{\nu }\eta ^{cd}A_{\nu }=e_{c}^{\mu }\eta
^{cd}A_{d}=e_{c}^{\mu }A^{c}  \tag{A5}
\end{equation}

Analogously, we may prove $A_\mu =g_{\mu \nu }A^\nu =<\frac \partial
{\partial x^\mu },\frac \partial {\partial x^\nu }>A^\nu =e_\mu ^aA_a.$
Thus, these studies are consistent.

Appendix B

Similar to deducing the important invariant $\omega _\mu \omega ^\mu $ of
inner product of two 1-forms, therefore, we deduce an important invariant of
inner product of two 2-form $\Omega $ under coordinate transformation, as
follows

\begin{equation*}
<<\Omega ,\Omega >>=<<\frac{1}{2}\Omega _{\mu \nu }dx^{\mu }\Lambda dx^{\nu
},\frac{1}{2}\Omega _{\mu ^{\prime }\nu ^{\prime }}dx^{\mu ^{\prime
}}\Lambda dx^{\nu ^{\prime }}>>
\end{equation*}%
\begin{equation}
=\frac{1}{4}\Omega _{\mu \nu }\Omega _{\mu ^{\prime }\nu ^{\prime }}g^{\nu
\mu ^{\prime }}<dx^{\mu },dx^{\nu ^{\prime }}>=-\frac{1}{4}\Omega _{\mu \nu
}\Omega ^{\mu \nu },  \tag{B1}
\end{equation}%
substituting Eq.(5) into the deduced inner product expression (B1) and
taking trace, then we have

\begin{equation}
tr<<\Omega _{V},\Omega _{V}>>=-\frac{1}{4}tr(\Omega _{V\mu \nu }\Omega
_{V}^{\mu \nu })=tr<<\Omega _{U},\Omega _{U}>>=-\frac{1}{4}tr(\Omega _{U\mu
\nu }\Omega _{U}^{\mu \nu }),  \tag{B2}
\end{equation}%
which satisfies gauge invariant principle \cite{wf} of general gauge field's
invariant, further using the invariant volume elements $(-g_{U})^{\frac{1}{2}%
}d\tau _{U}=(-g_{V})^{\frac{1}{2}}d\tau _{V}$ \cite{hou}$,$ we finally
achieve the invariant action Eq.(10) of general gauge fields in the global
curved spacetime.

\end{document}